\documentclass[10pt,conference]{IEEEtran}
\usepackage[paper=letterpaper,centering,margin=.71in]{geometry}
\usepackage{cite}
\usepackage{texgraphicx}
\graphicspath{{fig/}}
\usepackage{ifthen}
\usepackage[T1]{fontenc}
\usepackage{enumerate}
\usepackage[american]{babel}
\usepackage{amsfonts,amsmath,amssymb,amstext,latexsym,stmaryrd}
\usepackage{xspace}
\usepackage{subfigure}

\usepackage{amsthm_alvin}
\renewcommand{\qedsymbol}{\rule{3mm}{3mm}}
\theoremstyle{remark}

\newtheorem{theorem}{Theorem}
\newtheorem{proposition}{Proposition}
\newtheorem{corollary}{Corollary}
\newtheorem{definition}{Definition}
\newtheorem{remark}{Remark}

\makeatletter
\newcommand{\eqnlabel}[1]{\xdef\@currentlabel{\theequation}\ltx@label{#1}}
\newcommand{\Cnstslabel}[2][\thesubcounstraints]{
  ~\refstepcounter{subcounstraints}%
   \xdef\@currentlabel{#1}%
   \tagform@{#1}%
   \ltx@label{#2}}

\newcommand{\cnstlabelBIS}[2][\thesubcounstraints]{
   \xdef\@currentlabel{#1}%
   \ltx@label{#2}}
\newcounter{subcounstraints}[equation]

\newenvironment{LinearProgram}[2][Maximize]{%
   \begin{equation*}%
     \let\label=\eqnlabel
     \begin{array}{l}%
       \textsc{#1 } #2 %
       \textsc{ under the constraints}\\%
       \begin{cases}%
}{%
       \end{cases}
     \end{array}
   \end{equation*}%
}

\newcommand{\EquationsNumbered}[1]{%
   \setcounter{subcounstraints}{0}
   \renewcommand{\thesubcounstraints}{\theequation\alph{subcounstraints}}
   \def\set@counstraintscounter{
    ~\refstepcounter{subcounstraints}%
     \xdef\@currentlabel{\thesubcounstraints}%
     \tagform@{\thesubcounstraints}%
   }
   \def\mark{\set@counstraintscounter\quad&}
   \def\n{\\\mark}
   \begin{aligned}%
     #1
   \end{aligned}%
}
\makeatother

\def\lb{\ensuremath{\llbracket}}  
\def\rb{\ensuremath{\rrbracket}}  
\let\leq=\leqslant
\let\geq=\geqslant
\let\epsilon=\varepsilon
\let\rho=\varrho

\def\mbold{\mathversion{bold}}

\def\ie{i.e.\xspace}
\def\eg{e.g.,\xspace}
\def\th{\ensuremath{{}^{th}}\xspace}
\newcommand{\calc}[1]{\ensuremath{w_{#1}}\xspace}

\newcommand{\com}[1]{\ensuremath{b_{#1}}\xspace}

\newcommand{\CALC}[1]{\ensuremath{W_{#1}}\xspace}
\newcommand{\ACALC}[1]{\ensuremath{\widetilde{W}_{#1}}\xspace}
\newcommand{\COM}[1]{\ensuremath{B_{#1}}\xspace}
\newcommand{\ACOM}[1]{\ensuremath{\widetilde{B}_{#1}}\xspace}
\newcommand{\Ncalc}[1]{\ensuremath{N^{(W)}_{#1}(t)}\xspace}
\newcommand{\Ncom}[1]{\ensuremath{N^{(B)}_{#1}(t)}\xspace}
\newcommand{\done}[2]{\ensuremath{\mathit{done}_{#1}(#2)}\xspace}
\newcommand{\cons}[3][]{\ensuremath{\alpha\ifthenelse{\equal{#1}{}}{}{^{(#1)}}_{%
      \ifthenelse{\equal{#2}{} \AND \equal{#3}{}}{}{%
        \ifthenelse{\equal{#2}{}}{}{#2,}\ifthenelse{\equal{#3}{}}{*}{#3}}}}\xspace}
\newcommand{\Acons}[3][]{\ensuremath{\widetilde{\alpha}\ifthenelse{\equal{#1}{}}{}{^{(#1)}}_{%
      \ifthenelse{\equal{#2}{} \AND \equal{#3}{}}{}{%
        \ifthenelse{\equal{#2}{}}{}{#2,}\ifthenelse{\equal{#3}{}}{*}{#3}}}}\xspace}

\newcommand{\consNC}[2][n]{\cons[nc]{#1}{#2}}

\newcommand{\AconsNC}[2][n]{\Acons[nc]{#1}{#2}}
\newcommand{\consC}[2][n]{\cons[\text{coop}]{#1}{#2}}
\newcommand{\consF}[2][n]{\cons[f]{#1}{#2}}
\newcommand{\consS}[2][n]{\cons[\Sigma]{#1}{#2}}

\newcommand{\TIME}[3][]{\ensuremath{\tau^{(B\ifthenelse{\equal{#1}{}}{}{,#1})}_{#2\ifthenelse{
        \NOT\equal{#2}{} \AND \NOT\equal{#3}{}}{,}{}#3}}\xspace}
\newcommand{\ATIME}[3][]{\ensuremath{\widetilde{\tau}^{(B\ifthenelse{\equal{#1}{}}{}{,#1})}_{#2\ifthenelse{
        \NOT\equal{#2}{} \AND \NOT\equal{#3}{}}{,}{}#3}}\xspace}
\newcommand{\Time}[2][n]{{\TIME[par]{#1}{#2}}{}\xspace}

\newcommand{\TimeS}[2][n]{{\TIME[seq]{#1}{#2}}{}}

\newcommand{\TIMECAL}[3][]{\ensuremath{\tau^{(W\ifthenelse{\equal{#1}{}}{}{,#1})}_{#2\ifthenelse{
        \NOT\equal{#2}{} \AND \NOT\equal{#3}{}}{,}{}#3}}\xspace}
\newcommand{\ATIMECAL}[3][]{\ensuremath{\widetilde{\tau}^{(W\ifthenelse{\equal{#1}{}}{}{,#1})}_{#2\ifthenelse{
        \NOT\equal{#2}{} \AND \NOT\equal{#3}{}}{,}{}#3}}\xspace}
\newcommand{\TimeCAL}[2][n]{{\TIMECAL[par]{#1}{#2}}{}\xspace}

\newcommand{\TimeSCAL}[2][n]{{\TIMECAL[seq]{#1}{#2}}{}}

\newcommand{\cc}[1][k]{\ensuremath{c_{#1}}\xspace}
\newcommand{\Cc}[1][n]{\ensuremath{C_{#1}}\xspace}

\newcommand{\W}[1][n]{\ensuremath{{\mathcal{W}_{#1}}}\xspace}

\newcommand{\B}[1][n]{\ensuremath{{\mathcal{B}_{#1}}}\xspace}

\def\satB{\ensuremath{\mathit{sat}\B}\xspace}
\def\satW{\ensuremath{\mathit{sat}\W}\xspace}

\begin{document}

\title{Non-Cooperative Scheduling of \\
  Multiple Bag-of-Task Applications}

\makeatletter
\let\thanks\@IEEESAVECMDthanks%
\makeatother

\author{\authorblockN{Arnaud Legrand and Corinne
    Touati$^*$\thanks{*The
      authors would like to thank the University of Tsukuba and the
      Japan Society for the Promotion of Science for supporting this
      work.}}
  \authorblockA{Laboratoire LIG, Grenoble, CNRS-INRIA, MESCAL project, France,\\
    \texttt{arnaud.legrand@imag.fr, corinne.touati@imag.fr} }
}

\maketitle

\begin{abstract}
  Multiple applications that execute concurrently on heterogeneous
  platforms compete for CPU and network resources. In this paper we
  analyze the behavior of $K$ non-cooperative schedulers using the
  optimal strategy that maximize their efficiency while fairness is
  ensured at a system level ignoring applications characteristics.  We
  limit our study to simple single-level master-worker platforms and
  to the case where each scheduler is in charge of a single
  application consisting of a large number of independent tasks. The
  tasks of a given application all have the same computation and
  communication requirements, but these requirements can vary from one
  application to another. In this context, we assume that each
  scheduler aims at maximizing its throughput. We give closed-form
  formula of the equilibrium reached by such a system and study its
  performance. We characterize the situations where this Nash
  equilibrium is optimal (in the Pareto sense) and show that even
  though no catastrophic situation (Braess-like paradox) can occur,
  such an equilibrium can be arbitrarily bad for any classical
  performance measure.
\end{abstract}


\section{Introduction}
\label{sec:intro}

The recent evolutions in computer networks technology, as well as
their diversification, yield to a tremendous change in the use of
these networks: applications and systems can now be designed at a much
larger scale than before. Large-scale distributed platforms (Grid
computing platforms, enterprise networks, peer-to-peer systems) result
from the collaboration of many people. Thus, the scaling evolution we
are facing is not only dealing with the amount of data and the number
of computers but also with the number of users and the diversity of
their needs and behaviors.  Therefore computation and communication
resources have to be \emph{efficiently} and \emph{fairly} shared
between users, otherwise users will leave the group and join another
one. However, the variety of user profiles requires resource sharing
to be ensured at a system level. We claim that even in a perfect
system where every application competing on a given resource receives
the same share and where no degradation of resource usage (\eg packet
loss or context switching overhead) occurs when a large number of
applications use a given resource, non-cooperative usage of the system
leads to important application performance degradation and resource
wasting.  In this context, we make the following contributions:

\begin{itemize}
\item We present a simple yet realistic situation where a fair and
  Pareto-optimal \emph{system-level} sharing fails to achieve an
  efficient \emph{application-level} sharing. More precisely, we study
  the situation where multiple applications consisting of large
  numbers of independent identical tasks execute concurrently on
  heterogeneous platforms and compete for CPU and network resources.
  SETI@home~\cite{SETI}, the Mersenne prime search~\cite{Prime},
  ClimatePrediction.NET~\cite{BOINC}, Einstein@Home~\cite{EINSTEIN},
  processing data of the Large Hadron Collider~\cite{LHC} are a few
  examples of such typical applications. As the tasks of a given
  application all have the same computation and communication
  requirements (but these requirements can vary for different
  applications), each scheduler aims at maximizing its throughput.
  This framework had previously been studied in a cooperative
  centralized framework~\cite{BCFLMR_IPDPS06}. In the previous
  context, at any instant, cooperation led to a dedication of
  resources to applications.  The system-level resource sharing aspect
  was therefore not present and is extensively described in
  Section~\ref{sec:model} of the present paper.

\item We characterize in Section~\ref{sec:multiport.1appli} the
  optimal selfish strategy for each scheduler (\ie the scheduling
  strategy that will maximize its own throughput in all circumstances
  and adapt to external usage of resources) and propose equivalent
  representations of such non-cooperative schedulers competition (see
  Section~\ref{sec:representations}).

\item The particularities of these representations enable us to
  characterize the structure of the resulting Nash equilibrium as well
  as closed-form values of the throughput of each application (see
  Section~\ref{sec:multiport.closedform}).

\item Using these closed-form formulas, we derive in
  Section~\ref{sec:pareto_opt} a necessary and sufficient condition on
  the system parameters (in term of bandwidth, CPU speed,~\dots) for
  the non-cooperative equilibrium to be Pareto-optimal.

\item We briefly study in Section~\ref{sec:ineff-measure} the
  well-known ``price of anarchy''~\cite{papa}. Unfortunately, this
  metric does not enable one to distinguish Pareto optimal points from
  non-Pareto optimal ones. That is why we propose an alternate
  definition, the ``\emph{selfishness degradation factor}''.

\item When studying properties of Nash equilibria, it is important to
  know whether paradoxical situations like the ones exhibited by
  Braess in his seminal work~\cite{braess} can occur. In such
  situations, the addition of resource (a new machine, more bandwidth
  or more computing power in our framework) can result in a
  simultaneous degradation of the performance of \emph{all} the users.
  Such situations only occur when the equilibrium is not
  Pareto-optimal, which may be the case in this framework. We
  investigate in Section~\ref{sec:braess} whether such situations can
  occur in our considered scenario and conclude with a negative
  answer.

\item Last, we show in Section~\ref{sec:performance}, that even when
  the non-cooperative equilibrium is Pareto-optimal, the throughput of
  each application is far from being monotonous with a resource
  increase. This enables us to prove that this equilibrium can be
  arbitrarily bad for any of the classical performance measures
  (average, maximal, and minimum throughput).
\end{itemize}

Section~\ref{sec:conclusion} concludes the paper with a discussion of
extensions of this work and future directions of research. Due to
space requirements, the proofs of the following theorems and
propositions will be omitted in this paper. The interested reader is
referred to~\cite{RR} for detailed proofs.

\section{Platform and Application Models}
\label{sec:model}

\subsection{Platform Model}
\label{sec:model.platform}

Our master-worker platform is made of $N+1$ processors
$P_0,P_1,\dots,P_N$. $P_0$ denotes the master processor, which does not
perform any computation. Each processor $P_n$ is characterized by its
computing power \CALC{n} (in $\text{Mflop.s}^{-1}$) and the capacity of
its connection with the master \COM{n} (in $\text{Mb.s}^{-1}$). Last,
we define the \emph{communication-to-computation ratio} \Cc of
processor $P_n$ as $\COM{n} / \CALC{n}$. This model leads us to the
following definition:
\begin{definition}
  We denote by \textbf{physical-system} a triplet $(N,B,W)$ where $N$
  is the number of machines, and $B$ and $W$ the vectors of size $N$
  containing the link capacities and the computational powers of the
  machines.
\end{definition}

\begin{figure}[htb]
  \centering
  \includegraphics[scale=.7]{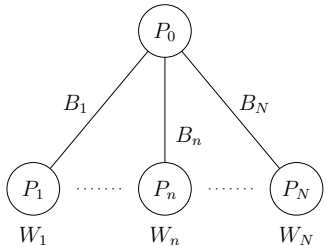}
  \caption{Platform model: a single master and $N$ workers}
\end{figure}

We assume that the platform performs an ideal fair sharing of
resources among the various requests. More precisely, let us denote by
\Ncom{n} (resp. \Ncalc{n}) the number of ongoing communication (resp.
computation) from $P_0$ to $P_n$ (resp. on $P_n$) at time $t$.
The platform ensures that the amount of bandwidth received at time $t$
by a communication from $P_0$ to $P_n$ is exactly $\COM{n}/\Ncom{n}$.
Likewise, the amount of processor power received at time $t$ by a
computation on $P_n$ is exactly $\CALC{n}/\Ncalc{n}$. Therefore, the
time $T$ needed to transfer a file of size \com{} from $P_0$ to $P_n$
starting at time $t_0$ is such that
\begin{equation*}
  \int_{t=t_0}^{t_0+T} \frac{\COM{n}}{\Ncom{n}}\cdot dt = \com{}.
\end{equation*}
Likewise, the time $T$ needed to perform a computation of size
\calc{} on $P_n$ starting at time $t_0$ is such that
\begin{equation*}
  \int_{t=t_0}^{t_0+T} \frac{\CALC{n}}{\Ncalc{n}}\cdot dt = \calc{}.
\end{equation*}
Last, we assume that communications to different processors do not
interfere. This amounts to say that the network card of $P_0$ has a
sufficiently large bandwidth not to be a bottleneck.  Therefore, a
process running on $P_0$ can communicate with as many processors as it
wants. This model is called
\emph{multi-port}~\cite{Banikazemi99,barnoy00multicast} and is
reasonable when workers are spread over the Internet, are not too
numerous and can make use of threads to manage
communications.

\subsection{Application Model}
\label{sec:model.application}

We consider $K$ applications, $A_k$, $1 \leq k \leq K$.  Each
application is composed of a large set of independent, same-size tasks.
We can think of each $A_k$ as a bag of tasks, in which each task is a file
that requires some processing.  A task of application $A_k$ is called a
task of \emph{type} $k$.  We let \calc{k} be the amount
of computation (in Mflop) required to process a
task of type $k$.  Similarly, \com{k} is the size (in Mb) of (the file
associated to) a task of type $k$.  We assume that the only
communication required is outwards from the master, \ie the
amount of data returned by the worker is negligible. This is a common
hypothesis~\cite{BBCFLR_TPDS04} as in steady-state, the output-file
problem can be reduced to an equivalent problem with bigger
input-files. Last, we define the \emph{communication-to-computation
ratio} \cc[k] of tasks of type $k$ as $\com{k} / \calc{k}$. This model
leads to the following definition:
\begin{definition}
  We define a \textbf{user-system} a triplet $(K,b,w)$ where $K$ is
  the number of applications, and $b$ and $w$ the vectors of size $K$
  representing the size and the amount of computation associated to
  the different applications.
\end{definition}

\subsection{Global System}
\label{sec:model.game}

In the following our $K$ applications run on our $N$ processors and
compete for network and CPU access:
\begin{definition}\label{def:system}
  A \textbf{system} $S$ is a sextuplet $(K,b,w,N,B,W)$, with
  $K$,$b$,$w$,$N$,$B$,$W$ defined as for a user-system and a
  physical-system.
\end{definition}

We assume that each application is scheduled by its own scheduler. As
each application comprises a very large number of independent tasks,
trying to optimize the makespan is known to be vainly
tedious~\cite{ejorDutot04} especially when resource availability
varies over time. Maximizing the throughput of a single application is
however known to be a much more relevant metric in our
context~\cite{BCFLMR_IPDPS06,bohong04}.  More formally, for a given
infinite schedule we can define \done{k}{t} the number of tasks of
type $k$ processed in time interval $[0,t]$.  The throughput for
application $k$ of such a schedule is defined as $\cons{}{k} =
\liminf_{t\to \infty} \frac{\done{k}{t}}{t}$.  Similarly we can define
\cons{n}{k} the average number of tasks of type $k$ performed per
time-unit on the processor $P_n$. \cons{}{k} and \cons{n}{k} are
linked by the following linear equation $\cons{}{k} = \sum_n
\cons{n}{k}$.  The scheduler of each application thus aims at
maximizing its own throughput, \ie \cons{}{k}.  However, as the
applications are sharing the same set of resources, we have the
following general constraints\footnote{The notation $\lb a,b\rb$
  denotes the set of integers comprised between $a$ and $b$, i.e.
  $\lb a,b\rb = \mathbb{N} \cap [a,b]$.}:
\begin{LaTeXdescription}
\item[Computation] $\forall n\in \lb 0,N\rb: \sum_{k=1}^K
  \cons{n}{k}\cdot\calc{k} \leq \CALC{n}$%
  \refstepcounter{equation}
  \renewcommand{\thesubcounstraints}{\theequation\alph{subcounstraints}}
  \hfill \Cnstslabel{eq.comp}
\item[Communication] $\forall n\in \lb 1,N\rb: \sum_{k=1}^K
  \cons{n}{k}\cdot\com{k} \leq \COM{n}$
  \hfill \Cnstslabel{eq.comm}
\end{LaTeXdescription}
These constraints enable to define upper-bounds on the throughput of
any schedule. Moreover, a \emph{periodic schedule} --- one that begins
and ends in exactly the same state --- (we refer the interested reader
to~\cite{j87} for more details) can be built from any valid
values for the \cons{}{k} and \cons{n}{k} such that its throughput for
all applications $k$ is exactly $\cons{}{k} = \lim_{t\to \infty}
\frac{\done{k}{t}}{t}$. When the number of tasks per application is
large, this approach has the advantage of avoiding the NP-completeness
of the makespan optimization problem. This further allows to only
focus on the average steady-state values.


\begin{remark} \label{rem:utility} Consider a system with $K$
  applications running over $N$ machines.  The set of achievable
  utilities, that is to say the set of possible throughput $\alpha_k$
  is given by
\begin{equation*}
  \scalebox{.9}{$
  U(S)=\left\{ (\cons{}{k})_{1 \leq k \leq K} \left|
      \begin{array}{@{\,}l@{}}
        \exists \cons{1}{1},\dots,\cons{N}{K}, \\
         \quad
        \begin{array}{l@{}}
          \forall k\in \lb 1,K\rb: \sum_{n=1}^{N} \cons{n}{k} = \cons{}{k} \\
          \forall n\in \lb 1,N\rb: \sum_{k=1}^K \cons{n}{k}\cdot\calc{k}
          \leq \CALC{n} \\
          \forall n\in \lb 1,N\rb: \sum_{k=1}^K \cons{n}{k}\cdot\com{k}
          \leq \COM{n}\\
          \forall n\in \lb 1,N\rb, \forall k\in \lb 1,K\rb: \cons{n}{k}
          \geq 0
        \end{array}
      \end{array}
    \right.
  \right\}\hspace{-3pt}.$}
\end{equation*}
The utility set is hence convex and compact.
\end{remark}

\subsection{A Non-Cooperative Game}
\label{sec:multiport.1appli}

We first study the situation where only one application is scheduled
on the platform. This will enable us to simply define the scheduling
strategy that will be used by each player (scheduler) in the more
general case where many applications are considered.  When there is
only one application, our problem reduces to the following linear
program:
\begin{LinearProgram}{\sum_{n=1}^N \cons{n}{1}}
  \EquationsNumbered{\mark
    \forall n\in \lb 0,N\rb: \cons{n}{1}\cdot\calc{1} \leq \CALC{n}\n%
    \forall n\in \lb 1,N\rb: \cons{n}{1}\cdot\com{1} \leq \COM{n}\n%
    \forall n, \quad \cons{n}{1} \geq  0.%
  }
\end{LinearProgram}
We can easily show that the optimal solution to this linear program is
obtained by setting $\forall n, \, \cons{n}{1} =
\min\left(\frac{\CALC{n}}{\calc{}}, \frac{\COM{n}}{\com{}}\right)$. In
a practical setting, this amounts to say that the master process will
saturate each worker by sending it as many tasks as
possible. On a stable platform \CALC{n} and \COM{n} can easily be
measured and the \cons{n}{1}'s can thus easily be computed. On an
unstable one this may be more tricky. However a simple acknowledgment
mechanism enables the master process to ensure that it is not
over-flooding the workers, while always converging to the optimal
throughput.

In a multiple-applications context, each player (process) strives to
optimize its own performance measure (considered here to be its
throughput $\cons{}{k}$) regardless of the strategies of the other
players. Hence, in this scenario, each process constantly floods the
workers while ensuring that all the tasks it sends are performed (\eg
using an acknowledgment mechanism). This adaptive strategy
automatically cope with other schedulers usage of resource and
\emph{selfishly} maximize the throughput of each application\footnote{We
suppose a purely non-cooperative game where no scheduler decides to
``ally'' to any other (\ie no coalition is formed).}. As the players
constantly adapt to each others' actions, they may (or not) reach some
equilibrium, known in game theory as \emph{Nash
  equilibrium}~\cite{NashEquilibrium50,NashEquilibrium51}. In the
remaining of this paper, we will denote by \consNC[n]{k} the rates
achieved at such stable states.

\subsection{A simple example}
\label{sec:multiport.example}

\begin{figure}[htb]
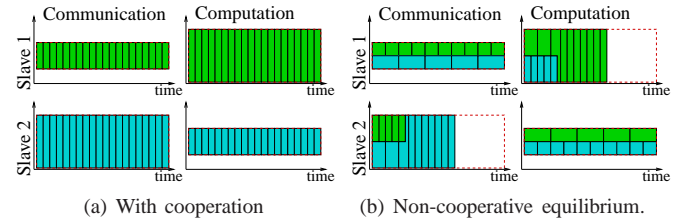

  \centering%
  \subfigure[With cooperation\label{fig:ex.1}]{
    \hspace{-.02\linewidth}%
    \includegraphics[width=.49\linewidth,subfig=2]{ineff2.fig}}%
  \subfigure[Non-cooperative equilibrium. \label{fig:ex.2}]{
    \hspace{-.01\linewidth}%
    \includegraphics[width=.49\linewidth,subfig=1]{ineff2.fig}}%
  \hspace{-2ex}%
  \caption{Non-cooperation can lead to inefficiencies.}
  \label{fig:ex}
\end{figure}

Consider a system with two computers $1$ and $2$, with parameters
$\COM{1} = 1$, $\CALC{1} = 2$, $\COM{2}= 2$, $\CALC{2} = 1$ and
two applications of parameters $\com{1} = 1$, $\calc{1} = 2$, $\com{2}
= 2$ and $\calc{2} = 1$.  If the applications were collaborating such
that application $1$ was processed exclusively to computer $1$ and
application $2$ in computer $2$ (see Figure~\ref{fig:ex.1}),
their respective throughput would be
\begin{equation*}
  \consC[]{1} = \consC[]{2} = 1.
\end{equation*}
Yet, with the non-cooperative approach (see
Figure~\ref{fig:ex.2}), one can check that they only get a
throughput of (the formal proof will be given in Theorem~\ref{th:NE}):
\begin{equation*}
  \consNC[]{1} = \consNC[]{2} = \frac{3}{4}
\end{equation*}

In this example, one can easily check that, at the Nash equilibrium,
for any worker, there is no resource waste: slave~1 (resp. slave~2) is
communication (resp. computation) saturated \ie
equation~\eqref{eq.comm} (resp. equation~\eqref{eq.comp}) is an
equality. However, communication-saturation implies computation idle
times and vice-versa. Yet, when communication and computation idle
times cohexist in a system, the non-cooperative behavior of the
applications can lead to important inefficiencies, which we
investigate further in the remaining of this paper.

\section{Mathematical Formulation}
\label{sec:representations}

In this section, we mathematically formulate the use of resources in
the system. As in the multi-port model communication resources are all
independent, we can study each worker separately.  Hence, in this
section, we study the use of resources on an arbitrary worker $n$ of
the system.

In a steady state, actions of the players will interfere on each
resource in (a priori) a non predictable order and the resource usage
may be arbitrarily complex (see Figure~\ref{fig:transfo.1}). We hence
propose in this section ``equivalent'' representations (in the sense
that they conserve the throughput of each application on the
considered worker) that will enable us to conclude this section with
a closed-form expression of the $(\alpha_{n,k})_{^{1 \leq n \leq N}_{1
    \leq k \leq K}}$.


First note that for any given subset $\mathcal{K}$ of $\lb 1,K \rb$,
we can define the fraction of time where all players of $\mathcal{K}$
(and only them) use a given resource.  This enables us to reorganize
the schedule into an equivalent representation (see
Figure~\ref{fig:transfo.2}) with at most $2^{|\mathcal{K}|}$
time intervals (the number of possible choices for the subset
$\mathcal{K}$). In this representation, the fractions of time spent
using a given resource (which can be a communication link or a
processor) are perfectly equal to the ones in the original schedule.
However such a representation is still too complex
($2^{|\mathcal{K}|}$ is a large value). Hence, we now explain
how to build two more compact ``equivalent'' canonical representations
(see Figure~\ref{fig:transfo.3} and~\ref{fig:transfo.4}).

\begin{figure*}[htb]
  \centering%
  \hspace{-.02\linewidth}%
  \subfigure[Complex arbitrary schedule\label{fig:transfo.1}]{
    \includegraphics[width=.03\linewidth,subfig=5]{schedule_transformation.fig}%
    \hspace{-.02\linewidth}%
    \includegraphics[width=.25\linewidth,subfig=1]{schedule_transformation.fig}}%
  \hspace{-2ex}%
  \subfigure[Sorted schedule\label{fig:transfo.2}]{
    \hspace{-.02\linewidth}%
    \includegraphics[width=.25\linewidth,subfig=2]{schedule_transformation.fig}}%
  \hspace{-2ex}%
  \subfigure[Sequential canonical representation: areas are preserved
  but using times are minimized\label{fig:transfo.3}]{
    \hspace{-.02\linewidth}%
    \includegraphics[width=.25\linewidth,subfig=4]{schedule_transformation.fig}}%
  \hspace{2ex}%
  \subfigure[Parallel canonical representation: areas are preserved
  but using times are maximized\label{fig:transfo.4}]{
    \hspace{-.02\linewidth}%
    \includegraphics[width=.25\linewidth,subfig=3]{schedule_transformation.fig}}%
  \caption{Various schedule representations. Each application is
    associated to a color: Application~1 is green, application~2 is
    yellow and application~3 is blue. The area associated to each
    application is preserved throughout all transformations.}
  \label{fig:transfo}
\end{figure*}

\subsection{Sequential Canonical Representation}
The first compact form we define is called \emph{sequential canonical
  representation} (see Figure~\ref{fig:transfo.3}). If the schedulers
were sending data one after the other on this link, the $k\th$
scheduler would have to communicate during exactly
$\TimeS{k}=\frac{\consNC{k} \com{k}}{\COM{n}}$ of the time to send the
same amount of data as in the original scheduler. This
value is called \emph{sequential communication time ratio}. Similarly,
we can define the \emph{sequential computation time ratio}
\TimeSCAL{k} as $\frac{\consNC{k} \calc{k}}{\CALC{n}}$. We hence have the
following relation between \TimeS{k} and \TimeSCAL{k}:
\begin{equation}
  \label{eq:mu_nu_relation}
  \TimeS{k} = \frac{\cc}{\Cc} \TimeSCAL{k}.
\end{equation}
We can therefore obtain a canonical schedule (see
Figure~\ref{fig:transfo.3}) with at most $K+1$
intervals whose respective sets of players are $\{1\}$, $\{2\}$,
$\dots$, $\{K\}$, $\emptyset$. This communication scheme is thus
called sequential canonical representation and has the same \consNC{k}
values as the original schedule. However, communication and computation
times have all been decreased as each scheduler is now using the
network link and the CPU exclusively. We will see later that this
information loss does not matter for multi-port schedulers.

\subsection{Parallel Canonical Representation}
The second compact form we define is called \emph{parallel canonical
  representation} (see Figure~\ref{fig:transfo.4}). In this scheme,
resource usage is as conflicting as possible. Let us denote by
$\Time{k}$ (resp. $\TimeCAL{k}$) the fraction of time spent by player
$k$ to communicate with $P_n$ (resp. to compute on $P_n$) in such a
configuration. \Time{k} is the \emph{parallel communication time
  ratio} and \TimeCAL{k} is the \emph{parallel computation time
  ratio}. We can easily prove that such representation is unique (see
the extended version~\cite{RR}) and we can therefore obtain a canonical
schedule (see Figure~\ref{fig:transfo.4}) with at most $K+1$ intervals
whose respective player sets are $\{1,\dots,K\}$, $\{2,\dots,K\}$,
$\dots$, $\{K\}$, and $\emptyset$.  This communication scheme is
called parallel canonical representation and has the same \consNC{k}
values as the original schedule. However, communication times have all
been increased as each scheduler is now interfering with as many other
schedulers as possible.

\subsection{Particularities of Multi-port Selfish Schedulers}
\begin{figure}[htb]
  \centering%
  \subfigure[Parallel canonical form of an arbitrary schedule
  \label{fig:canonical_schedule.1}]{
    \parbox[t]{.98\linewidth}{
      \includegraphics[scale=.5,subfig=1]{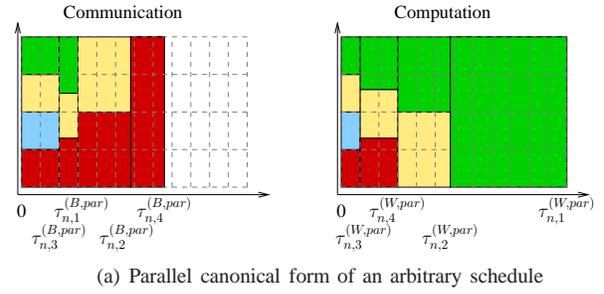}}}%

  \subfigure[Parallel canonical schedule for a given processor under
  the non-cooperative assumption. Application 3 (blue) and
  4 (red) are \emph{communication saturated}: they receive the same
  amount of bandwidth. Application 1 (green) and 2
  (yellow) are \emph{computation saturated}: they receive the same
  amount of CPU.
  \label{fig:canonical_schedule.2}]{
    \parbox[t]{.98\linewidth}{
      \includegraphics[scale=.5,subfig=1]{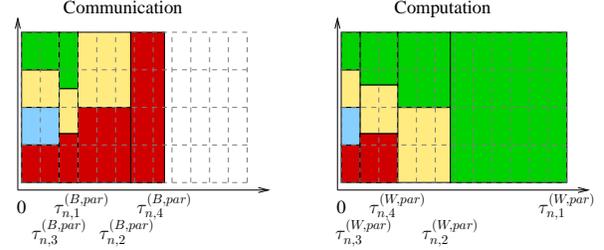}}%
  }
  \caption{Parallel canonical schedules}
  \vspace{-1.5em}
  \label{fig:canonical_schedule}
\end{figure}

The same reasonings can be applied to computation resources and
therefore, for a given worker, both communication and computation
resources can be put in any of these two canonical forms (see
Figure~\ref{fig:canonical_schedule.1}).

As we have seen in Section~\ref{sec:multiport.1appli}, the scheduling
algorithm used by the players consists in constantly flooding workers.
Hence it is impossible that both $\Time{k}$ and $\TimeCAL{k}$ are
smaller than 1. A player $k$ is thus said to be either 
\emph{communication-saturated} on worker $n$ ($\Time{k}=1$) or
\emph{computation-saturated} on worker $n$ ($\TimeCAL{k}=1$).

\begin{proposition}
  \label{prop:structure1}
  If there is a communication-saturated application then $\sum_{k=1}^K
  \TimeS{k} =1$. Similarly, if there is a computation-saturated
  application then $\sum_{k=1}^K \TimeSCAL{k} = 1$.
\end{proposition}

As two computation-saturated players $k_1$ and $k_2$ receive the same
amount of computation power and compute during the same amount of time,
we have $\consNC{k_1} \calc{k_1} = \consNC{k_2} \calc{k_2}$. Therefore
$\cc[k_1]\leq\cc[k_2]$ implies $\consNC{k_1}\com{k_1}\leq
\consNC{k_2}\com{k_2}$, hence $\Time{k_1} \leq \Time{k_2}$ and
$\TimeS{k_1} \leq \TimeS{k_2}$. The same reasoning holds for two
communication-saturated players as well as for a mixture of both. As a
consequence, in a multi-port setting, players should be first sorted
according to their \cc[k] to build the canonical schedule. The very
particular structure of this schedule (see
Figure~\ref{fig:canonical_schedule.2}) will enable us in the following
section to give closed-form formula for the \consNC[n]{k}. All these
remarks can be summarized in the following proposition:
\begin{proposition}
  \label{prop:structure2}
  Let us consider an equilibrium and denote by \B the set of
  communication-saturated applications on worker $n$ and by \W the set of
  computation-saturated applications on worker $n$.  If $\cc[1]\leq \cc[2] \leq
  \dots \leq \cc[K]$, then there exists $m\in \lb 0,K\rb$ such that
  $\W=\lb 1,m\rb$ and $\B=\lb m+1,K\rb$. We have:
  \begin{itemize}
  \item Sequential representation:
    Communications:
    \begin{equation*} 
      \scalebox{.9}{$
        \TimeS{1} \leq \dots \leq \TimeS{m}
        < \overbrace{\TimeS{m+1} = \dots = \TimeS{K}}^{\B} $
        < 1}
    \end{equation*}
    Computations:
    \begin{equation*} 
      \scalebox{.9}{$
        1 > \underbrace{\TimeSCAL{1}
          = \dots = \TimeSCAL{m}}_{\W} > \TimeSCAL{m+1} \geq \dots
        \geq \TimeSCAL{K}$
      }
    \end{equation*}
  \item Parallel representation:
    Communications: 
    \begin{equation*}
      \scalebox{.9}{$
        \Time{1} \leq \dots \leq \Time{m} <
        \overbrace{\Time{m+1} = \dots = \Time{K}}^{\B} = 1$
      }
    \end{equation*}
    Computations:
    \begin{equation*}
      \scalebox{.9}{$
        1 = \underbrace{\TimeCAL{1} = \dots =
          \TimeCAL{m}}_{\W} > \TimeCAL{m+1} \geq \dots \geq
        \TimeCAL{K}$
      }
    \end{equation*}
  \end{itemize}
\end{proposition}

\subsection{Closed-form Solution of the Equations}
\label{sec:multiport.closedform}

The closed-form solutions of the equilibrium are defined by
Theorem~\ref{th:NE}. Its complete proof, as well as the proofs of the
other propositions and theorems presented in the remaining of this
paper can be found in the extended version~\cite{RR}.

\begin{theorem} \label{th:NE}%
\label{sec:multiport.conditions}
We assume $\cc[1]\leq \cc[2] \leq \dots \leq \cc[K]$.  Let us denote
by \W the set of players that are computation-saturated and by \B the
set of players that are communication-saturated on a given arbitrary
worker $n$.
  \begin{enumerate}
  \item If $\sum_{k} \frac{\Cc}{\cc} \leq K $ then $\W=\emptyset$ and
    \begin{equation*}
      \displaystyle \forall k, \consNC{k} = \frac{\COM{n}}{K.\com{k}}.
    \end{equation*}
  \item Else, if $\sum_k \frac {\cc}{\Cc} \leq K$ then
    $\B=\emptyset$ and
    \begin{equation*}
      \displaystyle \forall k, \consNC{k} = \frac{\CALC{n}}{K.\calc{k}}.
    \end{equation*}
  \item Else, \B and \W are non-empty and there exists an integer $m \in \lb 1; K-1 \rb$ such that 
    \begin{equation*}
      \frac{\cc[m]}{\Cc} < \frac{m-\sum_{k=1}^m
        \frac{\cc}{\Cc}} {K-m - \sum_{k=m+1}^K \frac{\Cc}{\cc}} <
      \frac{\cc[m+1]}{\Cc}.
    \end{equation*}
    Then, we have $\W=\{1,\dots, m\}$ and $\B=\{m+1,\dots,K\}$ and
    \begin{equation}
      \label{eq.th:NE}
      \begin{cases}
      \consNC{k} = \frac {\COM{n}}{\com{k}} \frac{|\W|-\sum_{p \in \W}
        \frac{\cc[p]}{\Cc}}{|\W||\B| - \sum_{p \in \W} \cc[p] \sum_{p
          \in \B} \frac{1}{\cc[p]}} & \text{if }k \in \B \\
      \consNC{k} = \frac {\CALC{n}}{\calc{k}} \frac{|\B|- \sum_{p \in
          \B} \frac{\Cc}{\cc[p]}}{|\W||\B| - \sum_{p \in \W} \cc[p]
        \sum_{p \in \B} \frac{1}{\cc[p]}} & \text{if } k \in \W \\
      \end{cases}
    \end{equation}
  \end{enumerate}
\end{theorem}

\begin{proof}[Sketch of the proof]
  If $\B=\emptyset$, then all applications use the CPU of $P_n$ at any
  instant. Therefore they all receive the exact same amount of CPU, \ie
  $\CALC{n}/K$. Hence we have $\consNC{k} = \frac{\CALC{n}}{K.\calc{k}}$.
  Moreover, $\forall k, \TimeSCAL{k} = 1/K$ and from
  \eqref{eq:mu_nu_relation} we have $1\geq \sum_k \TimeS{k} = \sum_k
  \frac {\cc}{\Cc} \TimeSCAL{k} = \sum_k \frac {\cc}{K\Cc}$.  Hence
  $\sum_k \frac{\Cc}{\cc} \leq K$. The case
  $\W=\emptyset$ is similar.

  Let us now focus on the more interesting case where both
  $\B\neq\emptyset$ and $\W\neq\emptyset$. Using the definition of
  sequential communication and computation times, we have:
  \begin{equation}
    \label{eq.th:NE.1}
    \begin{cases}
      \sum_{p \in \B} \TimeS{p} + \sum_{p \in \W} \TimeS{p} =1\\
      \sum_{p \in \B} \TimeSCAL{p} + \sum_{p \in \W} \TimeSCAL{p} =1\\
    \end{cases}
  \end{equation}
  Two applications from \B communicate all the time. Therefore they
  send the exact same amount of data that we denote by $\TIME{\mathcal{B}}{}$:
  $\displaystyle \forall k \in \B, \consNC{k} \frac {\COM{n}}
  {\com{k}} = \TimeS{k} = \TIME{\mathcal{B}}{}.
  $  
  Similarly, we get
  $ \displaystyle \forall k \in \W, \consNC{k} \frac {\CALC{n}} {\calc{k}} = \TimeSCAL{k} =
  \TIMECAL{\mathcal{W}}{}
  $.
  From these relations and from \eqref{eq:mu_nu_relation},
  system~\eqref{eq.th:NE.1} can be written:
  \begin{equation*}
    \begin{cases}
      |\B| \TIME{\mathcal{B}}{} + \TIMECAL{\mathcal{W}}{} \sum_{p \in \W}
      \frac{\cc[p]}{\Cc} = 1\\
      |\W| \TIMECAL{\mathcal{W}}{} + \TIME{\mathcal{B}}{} \sum_{p \in \B} \frac{\Cc}{\cc[p]}  = 1
    \end{cases}
  \end{equation*}
  which can be easily solved to get \eqref{eq.th:NE}.

  Let $m$ such that $m \in \B$ and $m+1 \in \W$. From ~\eqref{eq:mu_nu_relation} and \eqref{eq.th:NE} and Proposition~\ref{prop:structure2},
  we can write:
    \begin{equation*}
      \frac{\cc[m+1]}{\Cc} = \frac{\TimeS{m+1}}{\TimeSCAL{m+1}} >
      \frac{\TimeS{m+1}}{\TimeSCAL{m}} = \frac{m-\sum_{k=1}^m
        \frac{\cc}{\Cc}} {K-m - \sum_{k=m+1}^K \frac{\Cc}{\cc}}
    \end{equation*}
    \begin{equation*}
      \text{and }
      \frac{\cc[m]}{\Cc} = \frac{\TimeS{m}}{\TimeSCAL{m}} <
      \frac{\TimeS{m+1}}{\TimeSCAL{m}} = \frac{m-\sum_{k=1}^m
        \frac{\cc}{\Cc}} {K-m - \sum_{k=m+1}^K \frac{\Cc}{\cc}}
    \end{equation*}
    which leads to the condition on $m$.
    The reciprocity of the conditions on the sets relies on the
    application of the following technical result with $\gamma_k =
    \cc/\Cc$.

    Let $\gamma_1<\dots <\gamma_K$ be $K$ positive numbers.
    We have:
    \begin{enumerate}
    \item If $\sum_k 1/\gamma_k \leq K$ then $\sum_k \gamma_k >K$;
    \item If $\sum_k \gamma_k \leq K$ then $\sum_k 1/\gamma_k >K$;
    \item If $\sum_k \gamma_k >K$ and $\sum_k 1/\gamma_k >K$, then there
      exists exactly one $m\in \lb 1,K\rb$ such that:  
    \end{enumerate}
    \begin{equation*}
      \gamma_m < \frac{\sum_{k=1}^m 1-\gamma_k}{\sum_{k=m+1}^K
        1-\frac{1}{\gamma_k}} < \gamma_{m+1}. \qedhere
    \end{equation*}
\end{proof}

\begin{corollary}
  From these equations, we see that there always exists exactly one
  non-cooperative equilibrium.
\end{corollary}

\section{Inefficiencies and Paradoxes}
\label{sec:pareto_opt}
In this section, we study the inefficiencies of the Nash equilibria, in
the Pareto sense, and their consequences. Let us start by recalling
the definition of the Pareto optimality.
\def\S{\ensuremath{\mathcal{S}}\xspace}
\begin{definition}[Pareto optimality]
  Let $G$ be a game with $K$ players. Each of them is defined by a set
  of possible strategies $\S_k$ and utility functions $u_k$ defined on
  $\S_1 \times \cdots \times \S_K$.\footnote{Note that the utility of
    a player depends on its own strategy and on the strategies of all
    the other players.}
  A vector of strategy is said to be Pareto optimal if it is
  impossible to strictly increase the utility of a player without
  strictly decreasing the one of another. In other words,
  $(s_1,\dots,s_K) \in \S_1 \times \cdots \times \S_K$ is Pareto
  optimal if and only if:\\[-1.8\baselineskip]
  \begin{multline*}
    \forall (s^*_1,\dots,s^*_K) \in \S_1 \times \cdots \times \S_K,\\
      \exists i, u_i (s^*_1,\dots,s^*_K) > u_i (s_1,\dots,s_K)
      \Rightarrow \\
      \exists j, u_j (s^*_1,\dots,s^*_K) <
      u_j(s_1,\dots,s_K).
  \end{multline*}
\end{definition}

We recall that, in the considered system, the utility functions are the
$\cons{}{k}$, that is to say, the average number of tasks of
application $k$ processed per time-unit, while the strategies are the
scheduling algorithms (\ie which resources to use and when to use
them).

In this section, we comment on the efficiency of the Nash equilibrium,
in the case of a single worker (Section~\ref{sec:Pareto-single}), and
then of multiple workers (\ref{sec:Pareto-multiple}) and propose in
Section~\ref{sec:ineff-measure} a brief study of the well-known
``price of anarchy''.  Unfortunately, this metric does not enable to
distinguish Pareto optimal points from non-Pareto optimal ones. That
is why we also propose an alternate definition, the ``\emph{selfishness
degradation factor}''.
Last, it is known that when Nash equilibria are inefficient, some
paradoxical phenomenon can occur (see, for instance \cite{braess}). We
hence study in Section~\ref{sec:braess}, the occurrence of Braess
paradox in this system.

\subsection{Single Processor} \label{sec:Pareto-single}

We can show that when the players (here the applications) compete in a
non-cooperative way over a single processor, the resulting Nash
equilibrium is Pareto optimal (see the extended version~\cite{RR} for
a detailed proof).

\begin{proposition} \label{th:single}%
  On a single-processor system, the allocation at the Nash equilibrium
  is Pareto optimal.
\end{proposition}

\subsection{Multi-processors and Inefficiencies} \label{sec:Pareto-multiple}

Interestingly, although the Nash equilibria are Pareto optimal on any
single-worker system, we show in this section that these equilibria are not
always Pareto optimal for a system consisting of several processors.

We first exhibit this phenomenon on a simple example consisting of two
machines and two applications (Section~\ref{sec:ex-ineff}). We then
provide a very simple characterization of the systems under which the
non-cooperative competition leads to inefficiencies
(Section~\ref{sec:necessary}).

\subsubsection{Example of Pareto inefficiency} \label{sec:ex-ineff}
The Pareto optimality is a global notion. Hence, although for each
single-processor system, the allocation is Pareto optimal, the result
may not hold for an arbitrary number of machines. This phenomenon was
illustrated in Section~\ref{sec:multiport.example}.

\subsubsection{Necessary and sufficient condition} \label{sec:necessary}

We prove in~\cite{RR} the following very simple characterization of
the systems under which the non-cooperative competition leads to
inefficiencies.
\begin{theorem}
  Consider a system $S=(K,b,w,N,B,W)$ as defined in
  Definition~\ref{def:system}. Suppose that the applications are not all
  identical, that is to say that there exists $k_1$ and $k_2$ such
  that $\cc[k_1] < \cc[k_2]$. 
  
  Then, the allocation at the Nash equilibrium is Pareto inefficient if
  and only if there exists two workers, namely $n_1$ and $n_2$ such
  that $\W[n_1] = \emptyset$ and $\B[n_2] = \emptyset$.
\end{theorem}

\subsection{Measuring Pareto Inefficiency} \label{sec:ineff-measure}%
We have seen that the Nash equilibrium of the system can be Pareto
inefficient. A natural question is then ``how much inefficient is
it?''. Unfortunately, measuring Pareto inefficiency is still an open
question. It is hence the focus of this section.

\subsubsection{Definitions}
Papadimitriou~\cite{papa} introduced the now popular measure
``price of anarchy'' that we will study in this section.

Let us consider an efficiency measure $f$ on the \cons{}{k}.  For a
given system $S$ (\ie platform parameters along with the description
of our $K$ applications), we denote by $\consNC[]{k}(S)$, the rates
achieved on system $S$ by the non-cooperative algorithm. For any given
metric $f$, let $\left( \consF[]{k}(S) \right)_{1 \leq k \leq K}$ be a
vector of optimal rates on system $S$ for the metric $f$.  We define
the inefficiency $I_f(S)$ of the non-cooperative allocation for a
given metric and a given system as
\begin{equation*}
  I_f(S) = \frac{f\left(\consF[]{1}(S),\dots,\consF[]{K}(S)\right)}{
    f\left((\consNC[]{1}(S),\dots,\consNC[]{K}(S)\right)} \geq 1.
\end{equation*}
Papadimitriou focuses on the profit metric $\Sigma$ defined by
$\Sigma(\cons{}{1},\dots,\cons{}{K}) = \frac{1}{K}\sum_{k=1}^K
\cons{}{k}$. The price of anarchy $\phi_\Sigma$ is then be defined as
the largest inefficiency:
\begin{equation*}
  \phi_\Sigma = \max_S I_\Sigma(S) = \max_S
  \frac{\sum_k \consS[]{k}(S)}{
    \sum_k \consNC[]{k}(S)} \geq 1.
\end{equation*}

\subsubsection{Studying the Price of Anarchy on a simple example}
\def\SMK{\ensuremath{S_{M,K}}\xspace}
Let us consider the following simple system $\SMK$ defined by
$N=1$, \COM{1}=1, \CALC{1}=1, $\com{}=(\frac{1}{M},1,\dots,1)$, and
$\calc{}=(\frac{1}{M},1,\dots,1)$. It is then easy to compute the
following allocations (see Figure~\ref{fig:pareto}):
\begin{itemize}
\item $\cons[nc]{}{}(\SMK)=\left(\frac{M}{K},\frac{1}{K}, \dots,
    \frac{1}{K}\right)$ corresponds to the non-cooperative allocation;
\item $\cons[\Sigma]{}{}(\SMK)=\left(M, 0, \dots, 0\right)$ corresponds to
  the allocation optimizing the average throughput;
\item $\cons[\min]{}{}(\SMK)=\left(\frac{1}{K-1+1/M}, \dots,
    \frac{1}{K-1+1/M}\right)$ corresponds to the max-min fair
  allocation~\cite{equite};
\item $\cons[\Pi]{}{}(\SMK)=\left(\frac{M}{K},\frac{1}{K}, \dots,
    \frac{1}{K}\right)$ corresponds to the proportionally fair
  allocation which is a particular Nash Bargaining Solution~\cite{equite}.
  Surprisingly, on this instance, this allocation also corresponds to
  the non-cooperative one.
\end{itemize}
\begin{figure}
  \centering
  \includegraphics[width=.7\linewidth]{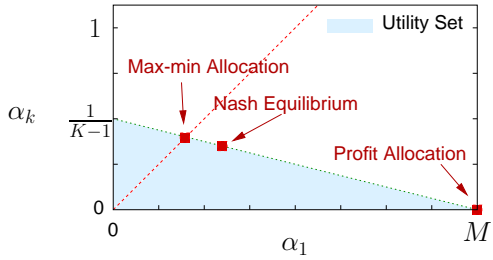}
  \caption{Utility set and allocations for $\SMK$ ($K=3$,$M=2$).}
  \label{fig:pareto}
\end{figure}
Note that, \cons[\Sigma]{}{}, \cons[\min]{}{}, and \cons[\Pi]{}{} are
Pareto optimal by definition. One can easily compute
\begin{equation*}
  I_\Sigma(\SMK) = \frac{M}{\frac{M}{K}+\frac{K-1}{K}}
  \xrightarrow[M\rightarrow\infty]{} K.
\end{equation*}
The price of anarchy is therefore unbounded. However, the fact that the
non-cooperative equilibria of such instances are Pareto-optimal and
have interesting properties of fairness (they correspond to a Nash
Bargaining Solution~\cite{equite}) questions the relevance of the
\emph{price of anarchy} notion as a Pareto efficiency measure.

Likewise, the inefficiency of the max-min fair allocation is
equivalent to $M$ for large values of $M$ (as opposed to $K$ for the
non-cooperative equilibrium). It can hence be unbounded even for
bounded number of applications and machines. This seems even more
surprising as such points generally result from complex cooperations
and are hence Pareto optimal. These remarks raise once more the
question of the measure of Pareto inefficiency.

\subsubsection{Selfishness Degradation Factor}

The previous problems are not specific to the efficiency measure
$\Sigma$. The same kind of behavior can be exhibited when using the
$\min$ or the product of the throughputs. That is why we think that
Pareto inefficiency should be measured as the \emph{distance} to the
Pareto border and not to a specific point.  

Based on the definition of the Pareto optimality, one can define the
concept of strict Pareto-superiority.
\begin{definition}[Pareto-superiority]
  A utility point $\alpha$ is said \emph{strictly Pareto-superior}
  to a point $\beta$ if for all player $k$ we have $\alpha_k > \beta_k$.
\end{definition}
Obviously, a Pareto-optimal point is such that there is no
achievable point strictly Pareto-superior to it.  To quantify the
degradation of Braess-like Paradoxes (the degree of Paradox),
Kameda~\cite{kameda06} introduced the Pareto-comparison of $\alpha$ and
$\beta$ as $\rho(\alpha,\beta) = \min_k \frac{\alpha_k}{\beta_k}$.
Therefore, $\alpha$ is strictly superior to $\beta$ iff
$\rho(\alpha,\beta)>1$.  Intuitively $\rho$ represents the performance
degradation between $\alpha$ and $\beta$. Using this definition,
we propose the following definition of Pareto inefficiency: $I(S) =
\max_{\alpha\in U(S)} \rho(\alpha,\cons[nc]{}{}(S))$.  Therefore
$\cons[nc]{}{}(S)$ is Pareto inefficient as soon as $I(S)>1$ and the
larger $\max_\alpha I(\alpha,\cons[nc]{}{})$, the more inefficient the
Nash equilibrium.  The \emph{selfishness degradation factor} can then
be defined as
\begin{equation*}
  \phi = \max_S I(S) = \max_S \max_{\alpha \in U(S)}
  \min_k \frac{\alpha_k}{\consNC{k}(S)}.
\end{equation*}
A system (\eg queuing network, transportation network, load-balancing,
...) that would be such that the Nash equilibria are always Pareto
optimal would have a selfishness degradation factor equal to one. The
selfishness degradation factor may however be unbounded on systems
where non-cooperative equilibria are particularly inefficient. The
relevance of this definition is corroborated by the fact that
$\epsilon$-approximations of Pareto-sets defined by Yannakakis and
Papadimitriou~\cite{papa_yannakakis} have a degradation factor of
$1+\epsilon$.  It can easily be shown that the systems studied in this
article have a selfishness degradation factor larger than two but the
exact value remains an open problem.

\subsection{Braess-like Paradoxes} \label{sec:braess}

When studying properties of Nash equilibria in routing systems, Braess
exhibited an example in which, by adding resource to the system (in
his example, a route), the performance of all the users were
degraded~\cite{braess}.  We investigate in this section whether such
situations can occur in our scenario.

Let us consider a system (called ``initial'') and a second one (referred
to as the ``augmented'' system), derived from the first one by adding
some quantity of resource. Intuitively, the Nash equilibrium $aug$ in
the augmented system should be Pareto-superior to the one in the
initial system $ini$. We say that a Braess paradox happens when $ini$
is strictly Pareto-superior to point $aug$.

Obviously, every achievable state in the initial system is also
achievable in the augmented system. Hence if $a$ is an achievable
point in the initial system and if $b$ is a \emph{Pareto} optimal
point is the augmented one, then $a$ cannot be strictly Pareto
superior to $b$. Hence Braess paradoxes are consequences of the Pareto
inefficiencies of the Nash equilibria.

We show that, even though the Nash equilibria may be Pareto
inefficient, in the considered scenario, Braess paradoxes cannot occur.

\begin{theorem} \label{th:braess}%
  In the non-cooperative multi-port scheduling problem, Braess like
  paradoxes cannot occur.
\end{theorem}

\begin{proof}[Sketch of the proof]
  We first need to introduce the definition of equivalent subsystem.
  Consider a system $S=(K,\com{},\calc{},N,\COM{},\CALC{})$. We define
  the new subsystem
  $\widetilde{S}=(K,\com{},\calc{},N,\ACOM{},\ACALC{})$ by: for each
  worker $n$,
  \begin{equation*}
\begin{array}{rl}
    &\ACALC{n} =
    \begin{cases}
      \sum_k \frac{\COM{n}}{K\cc} & \text{if $\W[n]=\emptyset$,}\\
      \CALC{n} & \text{otherwise,}
    \end{cases}\qquad\\\text{ and }
    &\ACOM{n} =
    \begin{cases}
      \sum_k \frac{\CALC{n}\cc}{K} & \text{if $\B[n]=\emptyset$,}\\
      \COM{n} & \text{otherwise.}
    \end{cases}
\end{array}
  \end{equation*}
  We now precise why $\widetilde{S}$ is said to be an equivalent
  subsystem of $S$. Consider a system
  $S=(K,\com{},\calc{},N,\COM{},\CALC{})$ and its Nash equilibrium
  $\consNC[]{}$. One can check that:
  \begin{enumerate}[i)]
  \item The system $\widetilde{S}$ is a subsystem of $S$, \ie for all
    worker $n$: $\ACOM{n}\leq \COM{n}$ and $\ACALC{n}\leq \CALC{n}$.
  \item The Nash equilibrium $\AconsNC[]{}$ of the subsystem
    $\widetilde{S}$ verifies
    \begin{equation*}
      \forall n, \forall k, \consNC{k} = \AconsNC{k}.
    \end{equation*}
  \item The Nash equilibrium $\AconsNC[]{}$ of the subsystem
    $\widetilde{S}$ is Pareto-optimal.
  \end{enumerate}
The conclusion of the proof relies on the following result:
  Consider two systems $S=(K,\com{},\calc{},N,\COM{},\CALC{})$ and
  $S'=(K,\com{},\calc{},N,\COM{}',\CALC{}')$ and their respective
  equivalent subsystems
  $\widetilde{S}=(K,\com{},\calc{},N,\ACOM{},\ACALC{})$ and
  $\widetilde{S}'=(K,\com{},\calc{},N,\ACOM{}',\ACALC{}')$. Suppose
  that $\forall n, \COM{n}' \geq \COM{n} \text{ and } \CALC{n}' \geq
  \CALC{n}$ then $\forall n, \ACOM{n}' \geq \ACOM{n} \text{ and }
  \ACALC{n}' \geq \ACALC{n}$.
\end{proof}

\section{Performance Measures}
\label{sec:performance}

In this section we show that unexpected behavior of some typical
performance measures can occur even for Pareto optimal situations. To ensure
optimality of the Nash equilibrium, we consider applications running
on a single processor (Proposition~\ref{th:single}).

We recall that the Pareto optimality is a global performance measure.
Hence, it is possible that, while the resources of the system increase
(either by the adding of capacity to a link or of computational
capabilities to a processor), a given performance measure decreases
while the equilibrium remains Pareto optimal.  The aim of this section
is to illustrate this phenomenon on some typical performance measures.

More precisely, we show the non-monotonicity of the
maximal throughput, of the minimal throughput and of the average
throughput. We finally end this section with a numerical example where
these measures decrease simultaneously with the increase
of the resource.


\subsection{Lower Bound on the Maximal Degradation} \label{sec:monot}

In the following, we suppose that only one of the resource of the
system increases. By symmetry, we suppose that the computational
capacity (\CALC{1}) is constant, while the link capacity \COM{1}
increases.\footnote{In the following, we will omit the subscript
  ``$1$'' as only one worker is considered in the system.}

Let us introduce $\underline{B} = \CALC{} K /\sum_{k} \frac{1}{\cc}$ and
$\overline{B} = \frac {\CALC{}}{K}\sum_k \cc$. From
Theorem~\ref{th:NE}, when considering the equations at the Nash
equilibrium, we can distinguish $3$ cases:
\begin{itemize}
\item 2 ``saturated'' situations that are:
  \begin{LaTeXdescription}
  \item[{\mbold \satW}] If $\COM{} \leq \underline{B}$, then
    $\consNC[]{k} = \frac{\COM{}}{K.\com{k}}$, \ie the throughput of
    each application is proportional to $\COM{}$. 
  \item[{\mbold \satB}] If $\COM{} \geq \overline{B}$ then
    $\consNC[]{k} = \frac{\CALC{}}{K.\calc{k}}$, \ie the throughput of
    each application is constant with respect with \COM{}.
  \end{LaTeXdescription}
\item 1 ``continuous'' situation when $\underline{B} <
  \COM{} < \overline{B}$
\end{itemize}

Obviously, in the ``saturated'' situations, the throughput
$\consNC[]{k}$ are increasing or constant and the order between the
applications is preserved. (I.e. if for $\COM{} \leq \underline{B}$
(resp. $\COM{} \geq \overline{B}$), $\consNC[]{k_1} \leq \consNC[]{k_2}$
then for all $\COM{}' \leq \underline{B}$ (resp.  $\COM{}' \geq
\overline{B}$) we have $\consNC[]{k_1} \leq \consNC[]{k_2}$.)

To simplify the analysis, we consider the degradation obtained when
$\COM{} = \underline{B}$ compared to the situation
where $\COM{} =\overline{B}$. It is hence a lower
bound on the actual maximum achievable degradation.

Consider an arbitrary application $k$. We write
${\consNC[]{k}}_\text{before}$ (resp. ${\consNC[]{k}}_\text{after}$) the
value of its throughput when $B=\underline{B}$ (resp.
$B=\overline{B}$). Hence, ${\consNC[]{k}}_\text{before} =
\frac{\COM{}}{K\com{k}}=\frac{\CALC{}}{\com{k} \sum_{p}
  \frac{1}{\cc[p]}}$ and ${\consNC[]{k}}_\text{after} =
\frac{\CALC{}}{K \calc{k}}$.  

\begin{remark}
  Note that $
  \frac{{\consNC{k}}_\text{before}}{{\consNC{k}}_{\text{after}}} =
  \frac{K}{\cc \sum_p 1/\cc[p]}$. The lower bound on the degradation
  is hence proportional to $1/\cc$ and is therefore maximal for the
  application with the smallest coefficient \cc. For instance, if
  $\forall p \neq k, \cc[p] = K$ and $\cc=1$, then
  $\frac{\alpha_\text{before}}{\alpha_{\text{after}}} \sim K/2$.
  Hence, when the number of applications grows to infinity, the
  degradation of the application having the smaller $\cc$ also grows
  to infinity.
\end{remark}

We can now easily show that even in a single processor system, the
maximal (and minimal) throughput can strictly decrease with the adding
of resource.
Note that:
  \begin{itemize}
  \item if $\COM{n} \leq \underline{B}$ the application $k$ having the
    highest (resp. smallest) throughput $\consNC{k}$ is the one having
    the smallest (resp. highest) value of $\com{k}$.
  \item if $B \geq \overline{B}$ the application $k$ having the
    highest (resp. smallest) throughput $\consNC{k}$ is the one whose
    $\calc{k}$ is the smallest (resp. highest).
  \end{itemize}
  Hence, a lower bound on the maximal degradation of the maximal
  (resp. minimal) throughput is $
  \frac {K}{\sum_k 1/\cc} \frac {\min_k \calc{k}} {\min_k \com{k}}$
  (resp. $
  \frac {K}{\sum_k 1/\cc} \frac {\max_k \calc{k}} {\max_k \com{k}}$ ).
  Therefore, for appropriate choices of $\min_k \com{k}$ and $\min_k
  \calc{k}$ (for all \cc fixed), the degradation can be chosen
  arbitrarily large.

  Finally, note that $\frac {\sum_k {\consNC{k}}_\text{before}}{\sum_k
    {\consNC{k}}_\text{after}} = \frac{K}{\sum_k 1/\cc} \cdot \frac
  {\sum_k 1/\com{k}}{\sum_k 1/\calc{k}}$. Hence, for some combinations
  of $\calc{k}$, $\com{k}$ and $\cc$, the degradation of the average
  throughput can be arbitrarily large (\eg for
  $\calc{}=(1,1/\epsilon,\dots,1/\epsilon)$ and
  $\com{}=(1,1/\epsilon^2,\dots,1/\epsilon^2)$, we get a ratio of
  $K\frac{1+(K-1)\epsilon^2}{(1+(K-1)\epsilon)^2}
  \xrightarrow[\epsilon \rightarrow 0]{} K$).


\subsection{Numerical Example}

We end this section with an example in which all the performance
measures we considered are simultaneously degraded when the bandwidth
\COM{} of the link connecting the master to the worker is increased.

\begin{figure}[htb]
  \centering
  \includegraphics[width=.8\linewidth]{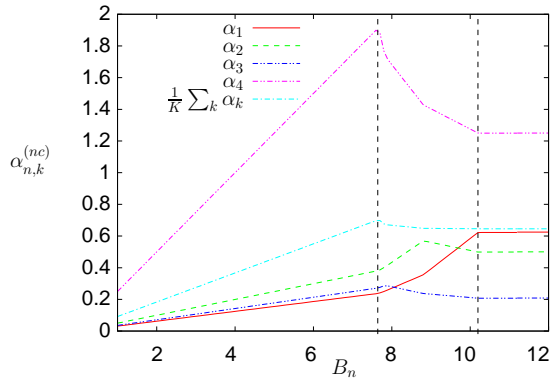}
  \vspace{-1em}
  \caption{The three performance measures can simultaneously decrease
    with the resource: \com{} = \{ 8,5,7,1\}, \calc{} = \{
    4,5,12,2\}, K=4, \CALC{}=10.}\label{fig:tous}
\end{figure}

Consider the example represented in Fig.~\ref{fig:tous}. Observe that
when the bandwidth $B$ is $245/24 \simeq 10.208$ the three measures
(namely the higher throughput, the lower throughput and the average
throughput) have lower values than when the bandwidth $B$ is only
equal to $560/73 \simeq 7.671$.

As $\min_k \calc{k} = \calc{4}$ and $\min_k \com{k} = \com{4}$, then
the application having the higher throughput in both \satB and \satW
is application $4$, and a lower bound of the degradation is $112/73$.

As $\max_k \calc{k} = \calc{3}$ and $\max \com{k} = \com{1}$, then the
application having the lower throughput is application $3$ in \satB
and application $1$ in \satW, and a lower bound of the degradation is
$84/73$.

Finally, a lower bound of the degradation of the average performance
is $2466/2263$.

\section{Conclusion}
\label{sec:conclusion}

We have presented a simple yet realistic situation where a fair and
Pareto-optimal \emph{system-level} sharing fails to achieve an
efficient \emph{application-level} sharing.  Even though the system
achieves a perfect sharing of resources between applications, the
non-cooperative usage of the system leads to important application
performance degradation and resource wasting. We have proved the
existence and uniqueness of the Nash equilibrium in our framework and
extensively studied its property. Surprisingly, the equilibrium is
Pareto-optimal on each worker independently. However, it may not be
globally Pareto-optimal. We have proved that no Braess-like
paradoxical situations could occur, which is, to the best of our
knowledge, the first situation where Pareto-inefficient
non-cooperative equilibrium cannot lead to Braess-like paradox.
However, some seemingly paradoxical situations can occur. Indeed, even
when the equilibria are Pareto optimal, their performance can be
arbitrarily bad for any classical performance measure.

This study led us to the natural question of the inefficency measure.
After briefly commenting on the notion of ``price of anarchy'', we
proposed a new definition, called SDF (Selfishness Degradation
Factor). 

The key hypothesis for deriving a closed-form description of the
equilibria is the multi-port hypothesis. Under this hypothesis, some
time information could be lost when using equivalent representations,
which resulted in simpler equations than if a 1-port model had been
used (i.e. if the master can communicate with only one worker at a
given instant). Preliminary simulations with this model show that
Braess-like paradoxes may occur. The understanding of such phenomena
are crucial to large-scale system planing and development as there is
no way to predict their apparition so far. Analytical
characterizations of such a framework could provide significant
insights on the key ingredients necessary to the occurrence of
Braess-like paradoxes.

Last, we can conclude from this study that cooperation between
applications is essential to avoid inefficiencies (even for simple
applications constituted of a huge number of independent identical
tasks). As far as the framework of this article is concerned, some
steps in this direction have been given in~\cite{BCFLMR_IPDPS06} where
some distributed algorithms were proposed and compared to an optimal
but centralized one. However, in their work, there was a single
scheduler whose duty was to achieve the best throughput for all
applications while ensuring a max-min fair share. In a
fully-decentralized setting, as considered in the present article,
some form of cooperation (\eg similar to the one proposed
by~\cite{Yaiche.YMR_TN00} for elastic traffic in broadband networks)
between different schedulers should be designed.

\bibliographystyle{IEEEtran}
\bibliography{biblio}

\end{document}